\documentclass[
superscriptaddress,aps,preprintnumbers,amsmath,amssymb,prd,nofootinbib,preprint]{revtex4-1}

\usepackage{graphicx}
\usepackage{epstopdf}
\usepackage{dcolumn}
\usepackage{bm}
\usepackage{hyperref}
\usepackage{color}
\usepackage{amsmath}
\usepackage{cancel}

\begin{document}


\def\a{\alpha}
\def\b{\beta}
\def\c{\varepsilon}
\def\d{\delta}
\def\e{\epsilon}
\def\f{\phi}
\def\g{\gamma}
\def\h{\theta}
\def\k{\kappa}
\def\l{\lambda}
\def\m{\mu}
\def\n{\nu}
\def\p{\psi}
\def\q{\partial}
\def\r{\rho}
\def\s{\sigma}
\def\t{\tau}
\def\u{\upsilon}
\def\v{\varphi}
\def\w{\omega}
\def\x{\xi}
\def\y{\eta}
\def\z{\zeta}
\def\D{\Delta}
\def\G{\Gamma}
\def\H{\Theta}
\def\L{\Lambda}
\def\F{\Phi}
\def\P{\Psi}
\def\S{\Sigma}

\def\o{\over}
\def\beq{\begin{align}}
\def\eeq{\end{align}}
\newcommand{\gsim}{ \mathop{}_{\textstyle \sim}^{\textstyle >} }
\newcommand{\lsim}{ \mathop{}_{\textstyle \sim}^{\textstyle <} }
\newcommand{\vev}[1]{ \left\langle {#1} \right\rangle }
\newcommand{\bra}[1]{ \langle {#1} | }
\newcommand{\ket}[1]{ | {#1} \rangle }
\newcommand{\EV}{ {\rm eV} }
\newcommand{\KEV}{ {\rm keV} }
\newcommand{\MEV}{ {\rm MeV} }
\newcommand{\GEV}{ {\rm GeV} }
\newcommand{\TEV}{ {\rm TeV} }
\newcommand{\1}{\mbox{1}\hspace{-0.25em}\mbox{l}}
\newcommand{\headline}[1]{\noindent{\bf #1}}
\def\diag{\mathop{\rm diag}\nolimits}
\def\Spin{\mathop{\rm Spin}}
\def\SO{\mathop{\rm SO}}
\def\O{\mathop{\rm O}}
\def\SU{\mathop{\rm SU}}
\def\U{\mathop{\rm U}}
\def\Sp{\mathop{\rm Sp}}
\def\SL{\mathop{\rm SL}}
\def\tr{\mathop{\rm tr}}
\def\mpl{M_{\rm Pl}}

\def\IJMP{Int.~J.~Mod.~Phys. }
\def\MPL{Mod.~Phys.~Lett. }
\def\NP{Nucl.~Phys. }
\def\PL{Phys.~Lett. }
\def\PR{Phys.~Rev. }
\def\PRL{Phys.~Rev.~Lett. }
\def\PTP{Prog.~Theor.~Phys. }
\def\ZP{Z.~Phys. }

\def\dd{\mathrm{d}}
\def\ff{\mathrm{f}}
\def\BH{{\rm BH}}
\def\inf{{\rm inf}}
\def\ev{{\rm evap}}
\def\eq{{\rm eq}}
\def\SM{{\rm sm}}
\def\Mpl{M_{\rm Pl}}
\def\GeV{{\rm GeV}}
\newcommand{\Red}[1]{\textcolor{red}{#1}}
\newcommand{\TL}[1]{\textcolor{blue}{\bf TL: #1}}


\title{
Nambu-Goldstone Affleck-Dine Baryogenesis
}

\author{Keisuke Harigaya}
\affiliation{School of Natural Sciences, Institute for Advanced Study, Princeton, NJ 08540, US}

\begin{abstract}
The Affleck-Dine mechanism creates the baryon asymmetry of the universe from an angular motion of a complex scalar field. The mechanism is efficient and allows for a low reheating temperature of the universe if the scalar field has a flat potential along its radial direction. We consider a possibility where the scalar field is a pseudo-Nambu-Goldstone boson arising from spontaneous breaking of an approximate global symmetry, so that the flatness of the potential is naturally understood. We construct a concrete realization of the idea based on chiral symmetry breaking in an $SU(N)$ gauge theory. The Peccei-Quinn mechanism can be consistently incorporated into the model. We also comment on a possible impact of the model on early universe physics.
\end{abstract}

\date{\today}

\maketitle

\section{Introduction}
Particle physics is aiming at understanding the origin of our universe. Any beyond the standard model physics should eventually explain how the baryon asymmetry of the universe has been created, namely, should contain a theory of baryogenesis.

Perhaps the most prominent mechanism is the leptogenesis~\cite{Fukugita:1986hr}; one of its predictions, non-zero masses of neutrinos, has been experimentally confirmed. The leptogenesis however requires that the reheating temperature of the universe is high enough~\cite{Giudice:2003jh,Buchmuller:2004nz}. Some scenarios of beyond the standard model are not compatible with high reheating temperatures because of overproduction of relics from the thermal bath~(e.g.~gravitinos~\cite{Pagels:1981ke,Weinberg:1982zq,Khlopov:1984pf,Ellis:1984er,Moroi:1993mb}), production of stable topological defects by low scale symmetry breaking after inflation (e.g.~the Twin Parity~\cite{Chacko:2005pe}, the left-right symmetry~\cite{Beg:1978mt,Mohapatra:1978fy,Kibble:1982ae,Chang:1983fu,Chang:1984uy,Babu:1989rb,Albaid:2015axa,Hall:2018let,Dunsky:2019api,Hall:2019qwx}, and the CP symmetry~\cite{Nelson:1983zb,Barr:1984qx,Bento:1991ez,Dine:2015jga}), etc.
Note that the maximal temperature of the universe is generically higher than the reheating temperature~\cite{Kolb:1990vq,Harigaya:2013vwa,Mukaida:2015ria} and hence the problem of production of stable topological defects is severe.
It is worth investigating theories of baryogenesis which do not require high reheating temperatures.

From this point of view, the Affleck-Dine (AD) baryogenesis~\cite{Affleck:1984fy} is attractive. Baryon asymmetry is created from an angular motion of a complex scalar field $\phi$ charged under an approximate $U(1)$ symmetry, which is called the AD field.
The resultant baryon asymmetry $n_B$ normalized by the entropy density $s$ is
\begin{align}
\frac{n_B}{s} \sim \epsilon \frac{\phi_i^2 T_{\rm RH}}{ m \mpl^2} = 10^{-10}\times \epsilon  \frac{10^{-9}}{m / \phi_i} \frac{T_{\rm RH}}{100~{\rm GeV}} \frac{\phi_i}{10^{16}{\rm GeV}},
\end{align}
where $\epsilon<1$ parametrizes the magnitude of explicit $U(1)$ breaking to drive the angular motion, $\phi_i$ is the initial field value of the radial direction of the AD field, $T_{\rm RH}$ is the reheating temperature, $\mpl$ is the reduced Planck scale, and $m$ is the mass of the radial direction around the field value $\phi_i$.
A larger reheating temperature leads to relatively larger asymmetry because of an earlier beginning of the red-shift of the radiation created by an inflaton.
The asymmetry increases as $m$ decreases since a later beginning of the oscillation of $\phi$ enhances the baryon number relative to that of radiation. To obtain enough baryon asymmetry for a low reheating temperature, the AD field must have a very flat potential such that $m \ll \phi_i$. A flat potential is natural in supersymmetric theories. In fact, sfermions in supersymmetric standard models have flat potentials and decay into baryons, and hence are natural candidates of the AD field~\cite{Affleck:1984fy,Dine:1995uk,Dine:1995kz}.

We consider an alternative scenario of the AD baryogenesis where the AD field is a pseudo-Nambu-Goldstone boson arising from spontaneous breaking of an approximate global symmetry.
We call the scenario as the Nambu-Goldstone Affleck-Dine (NGAD) baryogenesis. The idea is as follows. We assume that a global symmetry $G$ is spontaneously broken down to a subgroup $H$, yielding NG bosons on $G/H$. A $U(1)$ subgroup of $H$ is identified with the baryon symmetry. For an appropriate choice of $G$ and $H$, some of the NG bosons are charged under the $U(1)$ symmetry and is identified with the AD field $\phi$. (Part of) $G/H$ is explicitly broken but $U(1)$ conserving, which gives rise to the potential of the radial direction of $\phi$. As long as the explicit breaking is small, the required flatness of the potential of the AD field is guaranteed. Finally, the $U(1)$ symmetry is also explicitly broken by a small amount, driving angular motion of the AD field. The prescription is summarized as
\begin{gather}
{\rm Spontaneous}~~G \rightarrow H \supset U(1), \nonumber \\
{\rm Explicit}~~\cancel{G/H} \Rightarrow V_{U(1)}(\phi), \nonumber \\
{\rm Explicit}~~\cancel{U(1)} \Rightarrow V_{\cancel{U(1)}}(\phi). \nonumber 
\end{gather}
The example shown in the next section is analogous to a two flavor QCD with explicit breaking of global electromagnetic charge conservation; the analogue of the charged pion is the AD field. 

Symmetry breaking is ubiquitous in beyond the standard model and its origin may be unified with  that to yield the NGAD field. In fact, as is shown in the next section, it is possible to incorporate the Peccei-Quinn (PQ) mechanism~\cite{Peccei:1977hh,Peccei:1977ur} to the NGAD baryogensis. The model may then explain the baryon asymmetry of the universe, solve the strong CP problem~\cite{tHooft:1976rip}, and provide a QCD axion~\cite{Weinberg:1977ma,Wilczek:1977pj} as a dark matter candidate~\cite{Preskill:1982cy,Abbott:1982af,Dine:1982ah}.

\section{Affleck-Dine Baryogenesis by a Nambu-Goldstone boson}

\subsection{Model-independent analysis}

Let us first describe an idea of the NGAD baryogenesis in a model-independent way. We assume that a global symmetry $G$ is spontaneously broken down to a subgroup $H$, yielding NG bosons on $G/H$. A $U(1)$ subgroup of $H$ is identified with an approximate baryon or lepton symmetry, under which one of the NG bosons, $\phi$, is charged.
$G/H$ is explicitly broken and the NG boson $\phi$ obtains a $U(1)$ invariant potential
\begin{align}
V(\phi)_{U(1)} = m^2 f^2 J\left(\frac{|\phi|}{f}\right),
\end{align}
where $f$ is a decay constant and $m$ is the mass of $\phi$. $J$ is a function whose explicit form depends on the detail of the model.

We assume small explicit $U(1)$ symmetry breaking down to a $Z_n$ subgroup which drives angular motion of $\phi$ to generate $U(1)$ asymmetry,
\begin{align}
V(\phi)_{\cancel{U(1)}} =\epsilon m^2 f^2 \left[ \left( \frac{\phi}{f} \right)^n K\left(\frac{|\phi|}{f}\right) + {\rm h.c.} \right] + O(\frac{\phi^{2n}}{f^{2n}}),
\end{align}
where $\epsilon < 1$ parametrizes the small explicit breaking and $K$ is some function.

The field $\phi$ initially has a field value as large as $f$, and begins to oscillate when the Hubble expansion rate $H$ becomes smaller than the mass $m$. The asymmetry of $\phi$ given by
\begin{align}
n_\phi = 2{\rm Im} \left[ \phi^\dag \dot{\phi} \right]
\end{align}
obeys an equation of motion
\begin{align}
\dot{n_\phi} + 3H n_\phi  = {\rm Im} \left[ \phi \frac{\partial V }{ \partial \phi} \right].
\end{align}
The right hand side is proportional to $\epsilon$.
The asymmetry of $\phi$ created by the explicit $U(1)$ breaking per Hubble time is 
\begin{align}
\Delta n_\phi \sim \frac{\epsilon m^2 f^2}{H} \frac{\phi^n}{f^n} + {\rm h.c.} \sim \frac{\epsilon m^2 f^2}{H} \frac{H^n}{m^n},
\end{align}
where we assume a matter dominated era from which $\phi \propto H$ follows. Taking into account the red-shift, $\phi$ asymmetry is dominantly produced around the beginning of the oscillation if $n>3$, which we assume in the following.
The asymmetry of $\phi$ is then given by
\begin{align}
\frac{n_\phi}{s} \sim \frac{\epsilon f^2 T_{\rm RH} }{m \mpl^2},
\end{align}
where $T_{\rm RH}$ is the reheating temperature after inflation.

The asymmetry of $\phi$ is eventually transferred into baryon or lepton asymmetry. We consider the simplest possibility where this occurs via the decay of $\phi$ into standard model fermions with a decay rate $\Gamma_{\rm dec}$ around the temperature $T_{\rm dec} \sim \sqrt{\Gamma_{\rm dec} \mpl}$. Taking into account the possibility where the $\phi$ oscillation dominates the universe and creates entropy, the baryon or lepton asymmetry is given by
\begin{align}
\frac{n_{B,L}}{s} \sim {\rm min} \left[  \frac{\epsilon f^2 T_{\rm RH} }{m \mpl^2}, \frac{ \epsilon T_{\rm dec} }{m}  \right].
\end{align}
If lepton asymmetry is created before the electroweak phase transition, baryon asymmetry is created by the sphaleron process~\cite{Kuzmin:1985mm,Fukugita:1986hr,Harvey:1990qw}.

\subsection{A model based on an $SU(N)$ gauge theory}

\begin{table}[tb]
\caption{The matter content of a model of the NGAD baryogenesis.}
\begin{center}
\begin{tabular}{|c|c|c|c|c|}
\hline
               & $U$ & $D$ & $\bar{U}$ & $\bar{D}$ \\ \hline
$SU(N)$ & $N$ & $N$ & $\bar{N}$ & $\bar{N}$ \\
$U(1)$    & $1$ &  $-1$ & $-1$         & $1$ \\ \hline
\end{tabular}
\end{center}
\label{tab:charge}
\end{table}

We present a concrete model realizing the NGAD baryogenesis. We consider a model where the global symmetry $G$ is spontaneously broken by strong dynamics for the following reason. If  the symmetry breaking instead occurs by a fundamental scalar field, the natural value of the symmetry breaking scale is around the cut off scale. It is generically believed that global symmetry cannot be exact in theories with quantum gravity~\cite{Giddings:1988cx,Coleman:1988tj,Gilbert:1989nq,Banks:2010zn,Harlow:2018jwu,Harlow:2018tng}. Global symmetry would be then best understood as accidental symmetry, like the accidental baryon symmetry of the standard model. This idea will fail if the symmetry breaking scale is around the cut off scale, as any higher dimensional operators badly violate the global symmetry. We thus favor a model with a symmetry breaking scale much below the cutoff scale, which is naturally realized by dimensional transmutation from strong dynamics.

We introduce an $SU(N)$ gauge theory with two flavors.
The matter content of the model is shown in Table~\ref{tab:charge}. Below the dynamical scale, the global symmetry $G = SU(2)_L \times SU(2)_R$ is spontaneously broken down to $H= SU(2)_V$, yielding three NG bosons. The axial symmetry is explicitly broken, for example, by mass terms
\begin{align}
{\cal L} = m_U U\bar{U} + m_D D \bar{D}  + {\rm h.c.}~.
\end{align}
Up to this point, the theory possesses a $U(1)$ symmetry $\subset SU(2)_V$ shown in Table~\ref{tab:charge}. The NGAD field  $\phi$ is an analogue of the charge pion in the standard model.

The $U(1)$ symmetry is explicitly broken by the following interaction,
\begin{align}
{\cal L}= \frac{(4\pi)^{2n-2} c_1 }{M^{3n-4}}(U \bar{D})^n + {\rm h.c.}~,
\end{align}
where $M$ is the cutoff scale of the theory and $c_1$ is a constant. Here and hereafter, we fix the factor of $4\pi$s using the naive dimensional analysis (NDA)~\cite{Manohar:1983md,Georgi:1986kr,Luty:1997fk,Cohen:1997rt}. In this normalization, we require $c_1 <1$.
The parameter $\epsilon$ is as large as
\begin{align}
\epsilon \simeq \frac{c_1 (4\pi f)^{3n-2} }{m^2 M^{3n-4} },
\end{align}
where we use the NDA with a cut off scale at $4\pi f$.

The $U(1)$ charge retained in the angular motion of $\phi$ is eventually transferred into standard model fermions. We consider two possibilities.

\subsubsection{Decay into standard model fermions}
The simplest possibility is a decay of $\phi$ via a dimension-9 operator,
\begin{align}
{\cal L} = \frac{(4\pi)^4 c_2}{M^5} U \bar{D} \psi^4 + {\rm h.c.}~.
\end{align}
Here $\psi^4$ denotes an operator composed of four standard model fermions with a non-zero baryon or lepton charge, e.g.~$QQQL$, where $Q$ is a doublet quark and $L$ is a doublet lepton. As this is a very higher dimensional operator, the decay occurs before the Big-Bang-Nucleosynthesis only if the mass $m$ is large. Since the oscillation of the AD field occurs early, the produced baryon asymmetry cannot be as large as the observed one.

\subsubsection{Decay into a standard model fermion and a new heavy fermion}

The decay rate can be large enough if we introduce extra heavy fermions.
We consider the following dimension 6 operator,
\begin{align}
{\cal L} = \frac{(4\pi)^2 c_2}{M^2}U \bar{D} \psi \bar{\psi}' + {\rm h.c.}~,
\end{align}
where $\psi$ is a standard model fermion, while $\bar{\psi}'$ is a new heavy fermion which has a gauge charge opposite to $\psi$ and  a large Dirac mass term with another new fermion $\psi'$.  The $U(1)$ charges of $\psi$ and $\psi'$ are $-1$ and $1$, respectively.
The NGAD field $\phi$ decays into $\psi$ and $\bar{\psi}'$ with a rate
\begin{align}
\label{eq:decay}
\Gamma_{\rm dec} \simeq \frac{(4\pi)^5 c_2^2 f^4 m}{M^4},
\end{align}
and generates the asymmetry of  $\psi$ and $\psi'$.
Note that the parameter $c_2$ cannot be arbitrarily large, as quantum corrections generate an interaction
\begin{align}
\Delta{\cal L} \simeq \frac{c_2^2 (4\pi)^2}{ M^2} (U\bar{D})(U \bar{D})^\dag,
\end{align}
which gives a mass to $\phi$,
\begin{align}
\Delta m^2 \simeq \frac{(4 \pi)^4c_2^2 f^4}{M^2}.
\end{align}
We require that the correction is smaller than $m^2$.

The asymmetry of $\psi$ is transferred to baryon asymmetry and lepton asymmetry through the sphaleron process and the standard model yukawa interactions.
The new charged leptons $\psi'$ eventually decays into standard model particles by the following interaction which explicitly breaks the $U(1)$ symmetry,
\begin{align}
{\cal L} = y H \psi' \bar{\psi} + {\rm h.c.}~,
\end{align}
where $H$ is the standard model Higgs.
If $\psi'$ decays before the electroweak phase transition, net $\psi$ asymmetry vanishes and hence baryon and lepton asymmetry also vanish. The coupling $y$ must be sufficiently small so that $\psi'$ decays after the electroweak phase transition.

Let us see how baryon asymmetry survives in this case. To be concrete, let us consider the case where $\psi$ is a right-handed charged lepton $\bar{e}$. After $\phi$ decays, lepton asymmetry $L$ and $e'$ number asymmetry $E'$ are created,
\begin{align}
L = L_0,~
E' = - L_0.
\end{align}
The lepton asymmetry is partially converted into baryon asymmetry by the sphaleron process~\cite{Harvey:1990qw},
\begin{align}
B = - \frac{28}{79} L_0,~~
L = \frac{51}{79}L_0,~
E' = - L_0.
\end{align}
After the electroweak symmetry breaking, $B$ and $L$ are separately conserved. The asymmetry of $e'$ is converted into lepton asymmetry as $e'$ decays,
\begin{align}
B = - \frac{28}{79} L_0,~~
L = - \frac{28}{79}L_0.
\end{align}
Although the net $B-L$ asymmetry vanishes, the baryon and the lepton asymmetry are non-zero. In order for this to work, the decay of $\phi$ must occur before the electroweak symmetry breaking.

\begin{figure}
\centering
\includegraphics[width=0.47 \textwidth]{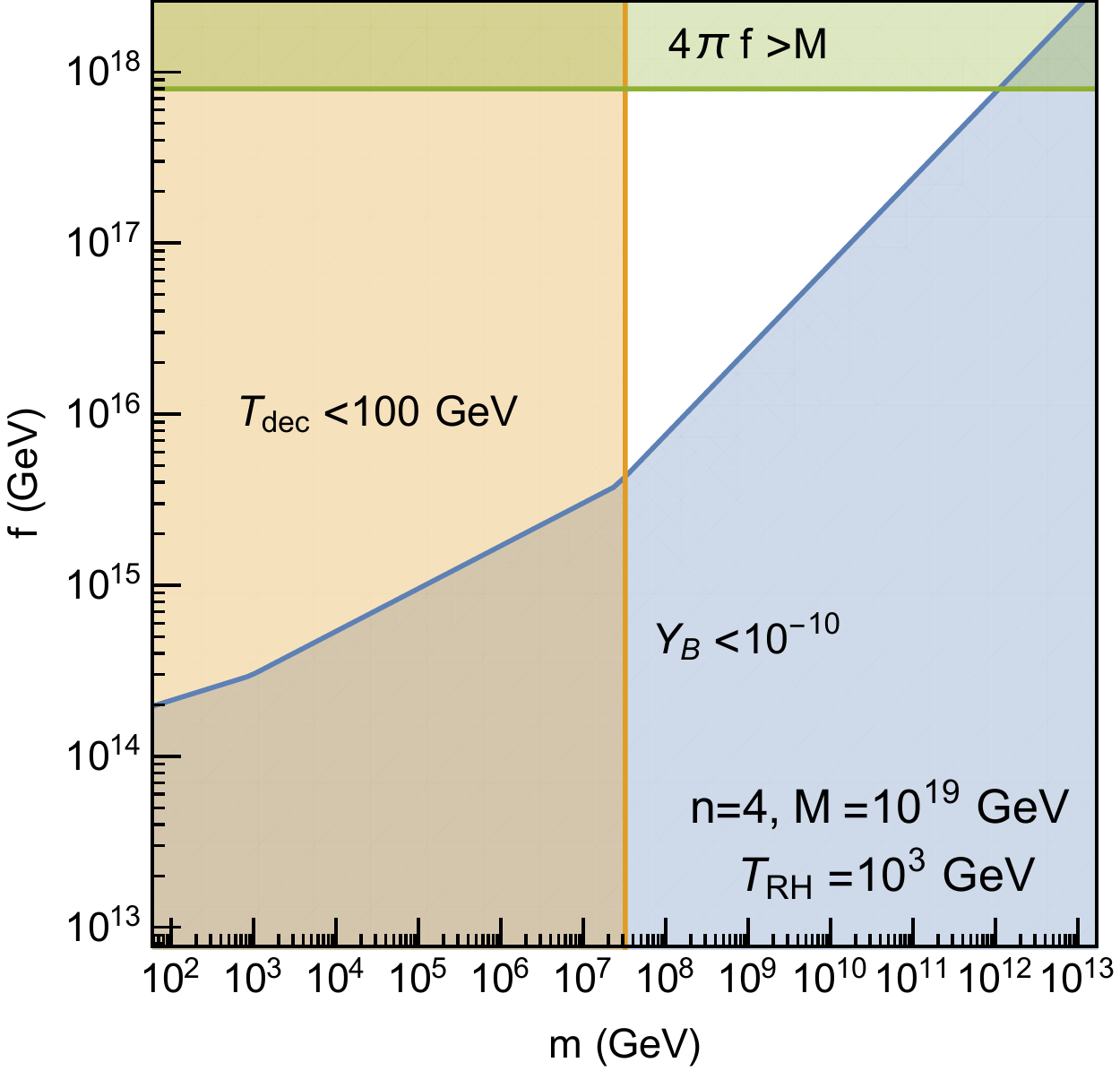} 
\includegraphics[width=0.47 \textwidth]{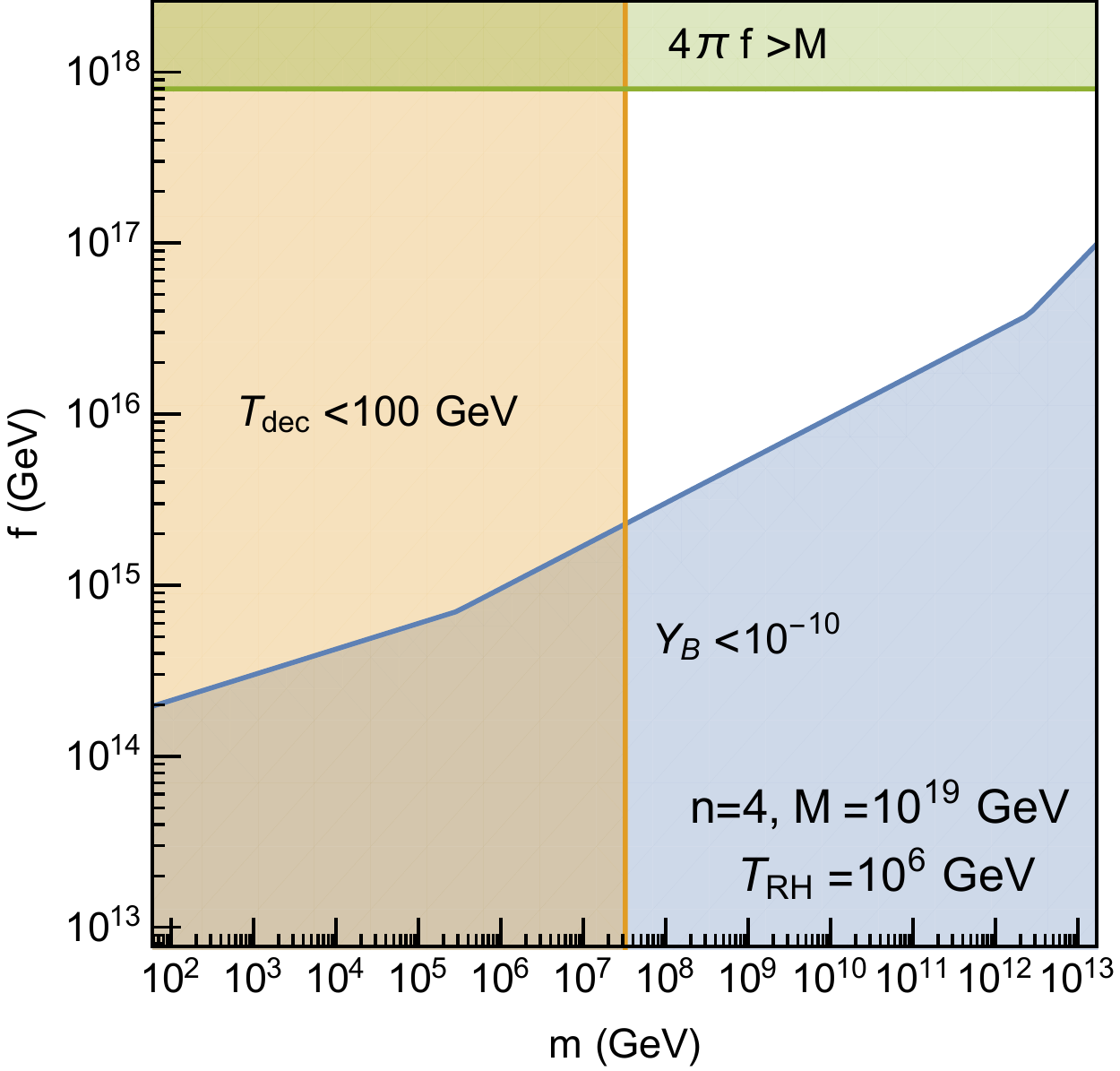}
\includegraphics[width=0.47 \textwidth]{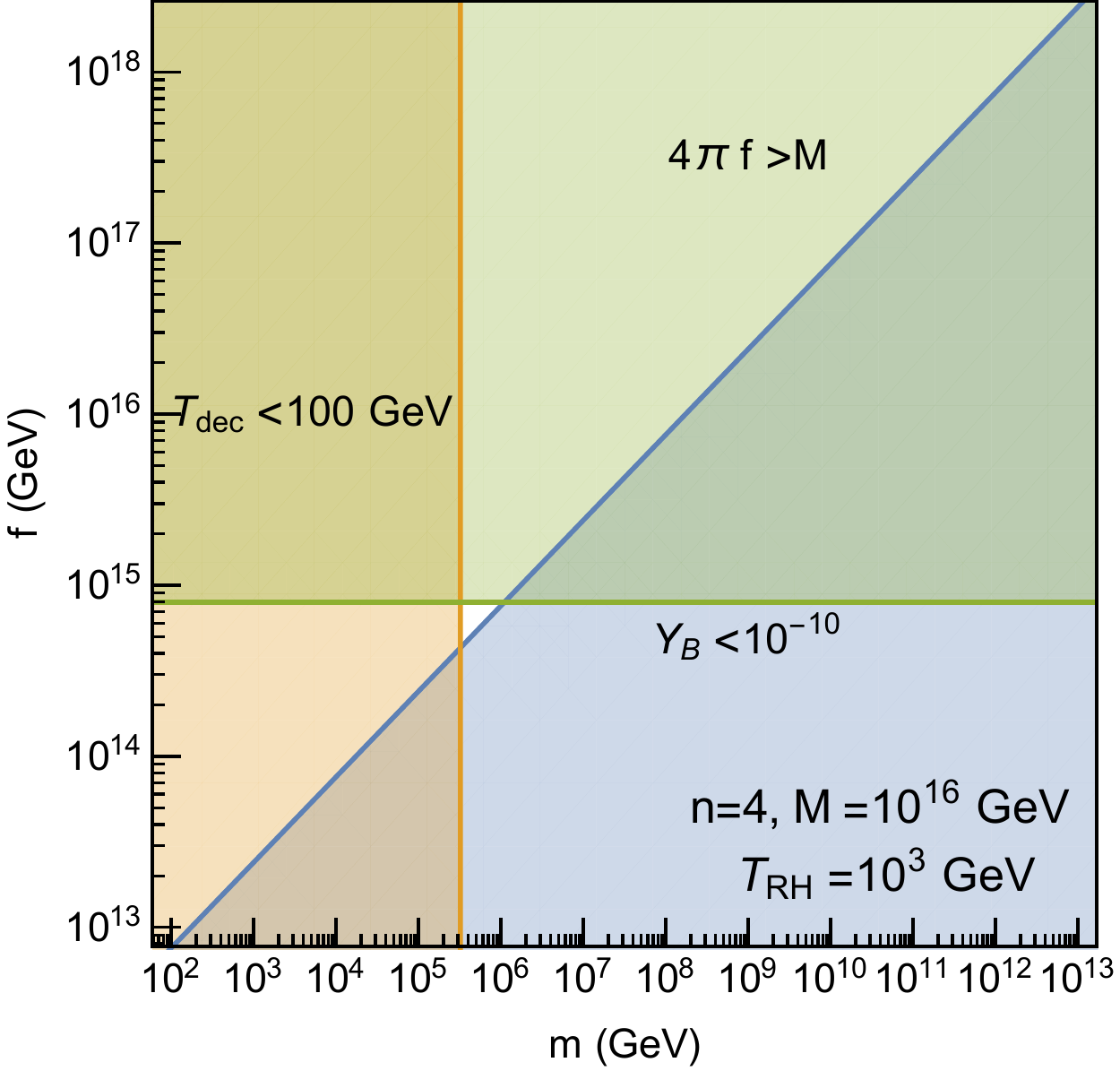} 
\includegraphics[width=0.47 \textwidth]{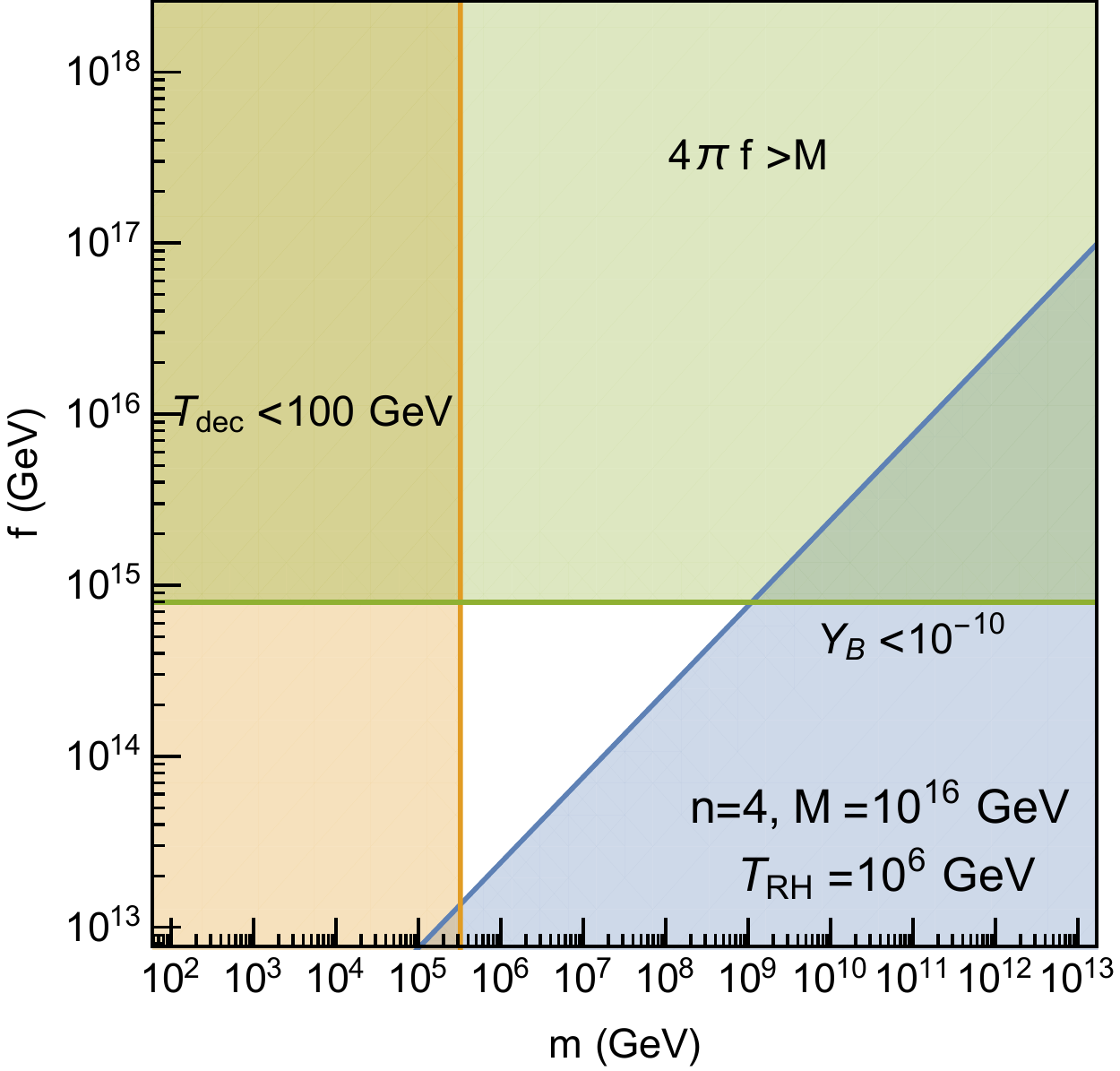} 
\caption{
Constraints on the mass $m$ and the decay constant $f$ of the NGAD field.
}
\label{fig:constraint}
\end{figure}

In Fig.~\ref{fig:constraint}, we show the bound on the mass $m$ and the decay constant $f$ for various reheating temperatures $T_{\rm RH}$ and cut off scales $M$.
In the blue-shaded region, the created baryon asymmetry is smaller than the observed one even if $c_1$ is as large as possible, satisfying $c_1 < 1$ and $\epsilon <1$. To the right of the rightmost kink in each plot, the constraint is determined by the condition $\epsilon <1$.
In the orange-shaded region, the decay of the AD field occurs after the electroweak phase transition even if $c_2$ is the maximal possible value satisfying $c_2<1$ and $\Delta m^2 < m^2$. The constraints shown in the figure are determined by the condition that $\Delta m^2 < m^2$. In the green-shaded region, the dynamical scale $4\pi f$ exceeds the cut off scale and the theory does not make sense. As the figure shows, even if the reheating temperature is as small as the weak scale, enough baryon asymmetry is produced. Also, in some of the allowed parameter space the symmetry breaking scale $\sim f$ is much below the cut off scale $M$. It may be possible to realize the global symmetry $G$ as an accidental symmetry arising from some exact symmetries. We leave the construction of such a model for future works.

We may consider a model without new standard model charged fermions but with a new pair of singlet Dirac fermions $S$ and $\bar{S}$. The interactions of $S$ and $\bar{S}$ with the NGAD field and standard model particles are
\begin{align}
{\cal L}= \frac{(4\pi)^2c_2}{M^2}U \bar{D} SS + y SLH + m_S S\bar{S} + {\rm h.c.}~.
\end{align}
The $U(1)$ charges of $S$ and $\bar{S}$ are $-1$ and $1$, respectively. In contrast to the model with new charged fermions, the yukawa coupling $y$ does not break the $U(1)$ symmetry. The AD field decays into $S$, which subsequently decays into $L$ and $H$ to create lepton asymmetry. In order for the sphaleron process to be efficient, both decays must occur before the electroweak symmetry breaking. The model with $S\bar{S}$ is subject to the same constraint on $(m,f)$ as the model with $\psi' \bar{\psi}'$. 

We comment on the effect of Q-ball formation. The radial direction of the AD filed is a pseudo-NG boson and has a cosine-potential with a period $\sim \pi f$.
The potential satisfies a criterion for the existence of a stable soliton called a Q-ball~\cite{Coleman:1985ki}, and we expect formation of Q-balls as the AD field starts oscillation~\cite{Kusenko:1997si,Enqvist:1997si,Enqvist:1998en,Kasuya:1999wu}. The typical radius and the charge of a Q-ball is about $m^{-1}$ and $ \epsilon f^2 / m^2$, respectively. The decay rate of such a Q-ball saturated by the Pauli-blocking effect is about $m^3/f^2/\epsilon $~\cite{Cohen:1986ct}. Because of the upper bound on $c_2$ from $\Delta m^2  < m^2$, the saturated decay rate is larger than the decay rate of a $\phi$ particle, and hence the decay rate of a Q-ball is simply given by Eq.~(\ref{eq:decay}).

\subsection{Inclusion of the Peccei-Quinn mechanism}

\begin{table}[tb]
\caption{The matter content of a model of the NGAD baryogenesis with a PQ symmetry.}
\begin{center}
\begin{tabular}{|c|c|c|c|c|c|c|}
\hline
               & $U$ & $D$ & $P$ & $\bar{U}$ & $\bar{D}$ & $\bar{P}$ \\ \hline
$SU(N)$ & $N$ & $N$ & $N$ & $\bar{N}$ & $\bar{N}$ & $\bar{N}$ \\
$SU(3)_c$ & $1$ & $1$ & $3$ & $1$ & $1$ & $\bar{3}$ \\
$U(1)_{\phi}$    & $1$ &  $-1$ & $0$ & $-1$ & $1$ & $0$ \\
$U(1)_{\rm PQ}$    & $0$ &  $-3$ & $1$ & $0$ & $0$ & $0$ \\ \hline
\end{tabular}
\end{center}
\label{tab:chargePQ}
\end{table}%

It is tempting to unify the origin of global symmetry breaking in the NGAD mechanism with other global symmetry breaking in beyond the standard model physics. 
Actually, we may easily incorporate the PQ mechanism~\cite{Peccei:1977hh,Peccei:1977ur} to the dynamics by adding massless $(SU(N),SU(3)_c)$ bi-fundamentals $P$ and $\bar{P}$. The matter content of the model and the PQ charge is shown in Table~\ref{tab:chargePQ}. The $U(1)$ symmetry of $\phi$ is denoted as $U(1)_\phi$. The Lagrangian presented above is invariant under the PQ symmetry if the mass $m_D = 0$. The PQ symmetry does not have an anomaly of $SU(N)$ while has that of $SU(3)_c$. In this model the PQ symmetry is spontaneously broken by strong dynamics, as in the model proposed in Ref.~\cite{Choi:1985cb}.

The chiral symmetry breaking pattern is $SU(5)_L \times SU(5)_R \rightarrow SU(5)_V$, yielding twenty four NG bosons. Twenty of them are $SU(3)_c$ charged, obeying $8$, $3$ and $3$ representations, and obtain large masses $\sim f$ by quantum corrections. Four of them are $SU(3)_c$ singlets. Among the four singlets, two form the AD field $\phi$, one obtains a mass from $m_U$, and one is massless in a PQ symmetric limit. The last one obtains a mass from the QCD strong dynamics and is a QCD axion, which solves the strong CP problem~\cite{tHooft:1976rip} and  is a dark matter candidate~\cite{Preskill:1982cy,Abbott:1982af,Dine:1982ah}.  

We simply impose the PQ symmetry as a global symmetry (which is actually not a symmetry as it is explicitly broken by the anomaly of $SU(3)_c$.) It will be interesting to construct a model where the PQ symmetry arises as an accidental symmetry as a result of other exact symmetry~\cite{Lazarides:1985bj,Holman:1992us,Barr:1992qq,Kamionkowski:1992mf,Dine:1992vx}.

\section{Summary and Discussion}

We have investigated a possibility of the Affleck-Dine baryogenesis by a Nambu-Goldstone boson. The flatness of the scalar potential of the AD field is naturally explained in this type of model without supersymmetry.
We have constructed a concrete example based on $SU(N)$ strong dynamics. We have found that enough baryon asymmetry can be produced even for a low reheating temperature.
It it possible to incorporate the Peccei-Quinn mechanism to the model.

We only consider the dynamics leading to baryogenesis in this paper. The rich structure of the model may further impact early universe physics. For example, since the radial direction of the AD field itself is a NG boson, it will be practically massless during inflation.
Then quantum fluctuations of the radial direction are produced during inflation~\cite{Mukhanov:1981xt,Hawking:1982cz,Starobinsky:1982ee,Guth:1982ec,Bardeen:1983qw}, which may source the primordial fluctuation of the universe. This is nothing but the curvaton scenario~\cite{Linde:1996gt,Enqvist:2001zp,Lyth:2001nq,Moroi:2001ct}. If the curvature perturbations are dominantly produced by the AD field, the upper bound on correlated baryon isocurvature perturbations~\cite{Akrami:2018odb} requires that the AD field decays after it dominates the energy density of the universe~\cite{Lyth:2002my} or that the baryon isocurvature perturbations are compensated by dark matter isocurvature perturbations~\cite{Holder:2009gd,Gordon:2009wx,Kawasaki:2011ze,Grin:2011tf,Grin:2013uya,Harigaya:2014bsa}. Other NG bosons in the model may also work as a curvaton. The phase direction of the AD field also obtains fluctuations and produces uncorrelated baryon isocurvature pertrubations of the same order of magnitude as the curvature perturbation, which is still consistent with the recent upper bound on them~\cite{Akrami:2018odb}.

\section*{Acknowledgement}
The author thanks Masahiro Ibe for useful discussion.
This work was supported in part by the Director, Office of Science, Office of High Energy and Nuclear Physics, of the US Department of Energy under Contracts DE-SC0009988.

\bibliography{draft}

\end{document}